\documentclass[10pt,a4paper,aps,showpacs,showkeys,
	amsmath,amssymb,pra,floatfix,
	preprint,
	superscriptaddress,
	longbibliography
	]{revtex4-1}

\usepackage{graphicx}
\usepackage{dcolumn}
\usepackage{bm}					
\usepackage{amsmath}
\usepackage{amssymb}

\usepackage[squaren]{SIunits}

\usepackage[usenames,dvipsnames]{xcolor}
\usepackage{slashed}
\usepackage{booktabs}					

\DeclareSymbolFont{rsfs}{U}{rsfs}{m}{n}
\DeclareSymbolFontAlphabet{\mathrsfs}{rsfs}
\usepackage[mathscr]{eucal}

\usepackage{hyperref}

\let\ud\d

\newcommand{\norm}[1]{||#1||}
\newcommand{\LCFAp}{LCFA\texttt{+}{ }}

\let\vec\boldsymbol

\begin{document}
	
\title{A Frenet-Serret Interpretation of Particle Dynamics in High-Intensity Laser Fields}

\author{D.~Seipt}
\email{dseipt@umich.edu}
\affiliation{Center for Ultrafast Optical Science, University of Michigan, Ann Arbor, Michigan 48109, USA}

\author{A.~G.~R.~Thomas}
\email{agrt@umich.edu}
\affiliation{Center for Ultrafast Optical Science, University of Michigan, Ann Arbor, Michigan 48109, USA}

\date{\today}

\begin{abstract}
	
In this paper we discuss the dynamics of charged particles in high-intensity laser fields in the context of the Frenet-Serret formalism,
which describes the intrinsic geometry of particle world-lines.
We find approximate relations for the Frenet-Serret scalars and basis vectors relevant for high-intensity laser particle interactions.
The onset of quantum effects relates to
the curvature radius of classical trajectories being on the order of the Compton wavelength.
The effects of classical radiation reaction are discussed, as
well as the classical precession of the spin-polarization vector according to the
{Thomas-Bargman-Michel-Telegdi (T-BMT)} equation.
We comment on the derivation of the photon emission rate in strong-field QED beyond the locally constant field approximation,
which is used in Monte Carlo simulations of quantum radiation reaction.
Such a numerical simulation is presented for a possible experiment to distinguish between classical and quantum mechanical
models of radiation reaction.
\end{abstract}

\maketitle

\section{Introduction}

The world lines of massive charged particles, such as electrons, are time-like curves through Minkowski space in the absence of gravitational fields. 
If forces are acting on the particles, those paths are curved. The internal geometry of such a world line can be described in an elegant way using the Frenet-Serret (FS) formalism
in terms of three scalar functions: The curvature $\kappa$ and two torsions, $\tau$ and $\sigma$, of the world line. Moreover, the FS formalism provides a tetrad of
orthogonal unit vectors, the tangent to the world line and three normals, that are transported along the world line by means of the Frenet-Serret equations \cite{Synge:ProcIrAcSc1967,Honig:JMathPhys1974,Formiga:AmJPhys2005,book:Synge}.
The FS formalism has also been applied to motion of charges in Riemann spaces, especially the ones with certain some symmetries admitting Killing vector fields \cite{Carmeli:PRB1965,Honig:JMathPhys1976}.
The relevance of symmetries for finding analytic solutions of classical and quantum motion of charges in background fields has been pointed out recently \cite{Heinzl:PRL2017}.

The interactions of charged particles with ultra-strong electromagnetic fields provided by high-intensity laser pulses allows to explore fundamental aspects of classical and quantum electrodynamics in extreme fields,
such as classical and quantum radiation reaction effects \cite{Cole:PRX2018,Poder:PRX2018}.
Of particular interest are nonlinear quantum effects governed by the parameter
{$\chi = \frac{e}{m^3} \sqrt{ p_\mu F^{\mu\nu}F_{\nu\alpha} p^\alpha}$}
which represents the laser electric field strength in the electron's rest frame in units of the critical Schwinger field {$E_S = m^2/e  \simeq \unit{1.32\times10^{18}}{\volt\per\metre}$}.
A better understanding of these effects will be necessary for 
the next generation of high-power lasers, reaching intensities of $10^{23}$ W/cm$^2$ and above
{where it is expected that $\chi\gtrsim1$ will be achieved routinely}, and their applications in laser-plasma based particle beam and photon sources {\cite{ELI-NP,Sung:OptLett2017,Kiriyama:OptLett2018,SEL100PW}.}

In this paper we will investigate the FS formalism to provide an alternative viewpoint
of the dynamics of charged particles in ultra-strong laser fields.
We find approximate relations for the Frenet-Serret scalars and the tetrad of basis vectors relevant for high-intensity laser particle interactions.
For instance, the onset of quantum effects for $\chi\sim 1$ relates to
the curvature radius of classical trajectories being on the order of the Compton wavelength.
We investigate the classical precession of the spin-polarization vector and focus on classical and quantum radiation reaction effects.
For the latter case we comment on the derivation of photon emission rates in strong-field QED beyond the locally constant field approximation.
Numerical simulations are presented showing the distinction of classical and quantum radiation reaction effects employing the improved photon emission rates.

We use natural Heaviside-Lorentz units with $\hbar=c=\epsilon_0=1$ and the fine structure constant $\alpha = e^2/4\pi$.
We will omit explicit notation of Lorentz indices whenever possible, i.e.~$u^\mu\to u$, $f^{\mu\nu}\to f$ and denote
scalar products and contractions of tensor indices as $u^\mu u_\mu \to u.u$, $f^{\mu\nu}f_{\nu\lambda} \to f.f = f^2$ etc.
Tetrad indices are denoted by uppercase Latin letters, e.g.~$c_A$.

\section{The Frenet-Serret Tetrad and the Frenet-Serret Equations}

The Frenet-Serret (FS) formalism in space-time defines a principal tetrad of orthonormal basis vectors $e_{(A)}^\mu$ which are transported along the world line $x^\mu(s)$ of a particle ($s$ denotes proper time), along with three associated scalars: 
the curvature $\kappa$ and two torsions $\tau$ and $\sigma$.
The orthonormality condition means that $e_{(A)}.e_{(B)} = g_{AB}$, where $g_{AB}$ denotes the components of the metric tensor.
The Frenet-Serret scalars describe the intrinsic geometry of the world-line and govern the transport of the tetrad along the world-line by means of the Frenet-Serret equations \cite{book:Synge}.

The first vector from the FS tetrad is the tangent to the world line $x(s)$, i.e.~the
four-velocity $e_{(0)} = u = \dot x$. It is a time-like unit-vector with $u.u =1$.
The space-like normal triad $\{e_{(1)},e_{(2)},e_{(3)}\}$ can be constructed from higher order derivatives of the world line using the Gram-Schmidt ortho-normalization procedure.
In particular, the normal vector $e_{(1)} = \frac{\dot u}{\norm{\dot u}}$, with the norm
$\norm{\dot u } = \sqrt{ - (\dot u.\dot u)} \equiv \kappa(s)$ defining the curvature $\kappa$ of the world line which is the magnitude of four-acceleration.
The binormal vectors $\{ e_{(2)} , e_{(3)}\}$ are constructed from higher proper-time derivatives of the world line,
assuming sufficient smoothness of the latter,
%
\begin{align*}
e_{(2)} &= \frac{E_2}{\norm{E_2}} \,, \qquad
E_2 = \ddot u - (\ddot u.e_{(0)}) e_{(0)} 
              + (\ddot u. e_{(1)}) e_{(1)} \,, \\
e_{(3)} &= \frac{E_3}{\norm{E_3}} \,, \qquad
E_3 = \dddot u - (\dddot u.e_{(0)}) e_{(0)} 
            + (\dddot u. e_{(1)}) e_{(1)} 
            +  (\dddot u. e_{(2)}) e_{(2)}  \,,
\end{align*}
and with the two torsions $\tau$ and $\sigma$ being defined as
$\kappa \tau =  \norm{E_2} $ and 
$\kappa\tau \sigma = \norm{E_3}$,
yielding the following useful expressions for the FS scalars
\begin{align}
\label{def:kappa}
\kappa^2                          &= -(\dot u. \dot u) \,,\ \\
\label{def:tau}
\kappa^2 \tau^2              &= -(\ddot u.\ddot u) + \kappa^4 
                                - \dot \kappa^2 \,, \\
\label{def:sigma}
\kappa^2\tau^2\sigma^2 &= -(\dddot u.\dddot u) - \kappa^2 (\kappa^2-\tau^2)^2 - \ddot \kappa^2
-2 \kappa^2 (\kappa^2-\tau^2) \frac{\ddot \kappa}{\kappa}
+ 9 \kappa^4 \frac{\dot \kappa^2}{\kappa^2}
-\kappa^2\tau^2 \left( \frac{\dot \tau}{\tau} + 2 \frac{\dot \kappa}{\kappa}\right)^2
\end{align}
This allows to write the FS tetrad, which is a complete orthonormal basis, as follows:
\begin{align}
e_{(0)} &= u \,, \qquad 
e_{(1)}  = \frac{ \dot u }{\kappa}\,, \qquad 
e_{(2)} =  \frac{1}{\kappa\tau} \left[ \ddot u - \kappa^2 u - \frac{\dot \kappa}{\kappa} \dot u \right] \,,\\
e_{(3)} &= 
\frac{1}{\kappa\tau\sigma} \left[ 
\dddot u 
- \left( \frac{\dot \tau }{\tau} + 2 \frac{\dot\kappa}{\kappa} \right) \,  \ddot u 
+ \left( \tau^2 -\kappa^2 
- \frac{\ddot \kappa}{\kappa} 
+ \frac{\dot\kappa}{\kappa} \frac{\dot \tau}{\tau}
+ 2\frac{\dot \kappa^2}{\kappa^2} \right) \dot u
+ \kappa^2 \left(
\frac{\dot \tau}{\tau} - \frac{ \dot \kappa }{\kappa}
\right) u
\right]	 \,.
\end{align}
In case that the second torsion vanishes, $\sigma=0$, the above definition of $e_{(3)}$ becomes singular.
Instead, one should use the alternative definition by means of the Levi-Civita tensor
\begin{align}
e_{(3)}^\mu &= \varepsilon^{\alpha\beta\gamma \mu } 
        e_{(0),\alpha} 
        e_{(1),\beta} 
        e_{(2),\gamma} = \frac{1}{\kappa^2\tau } \varepsilon^{\alpha\beta\gamma \mu } u_\alpha \dot u_\beta \ddot u_\gamma \,.
\end{align}

\begin{figure}
	\begin{center}
		\includegraphics[width=0.6\columnwidth]{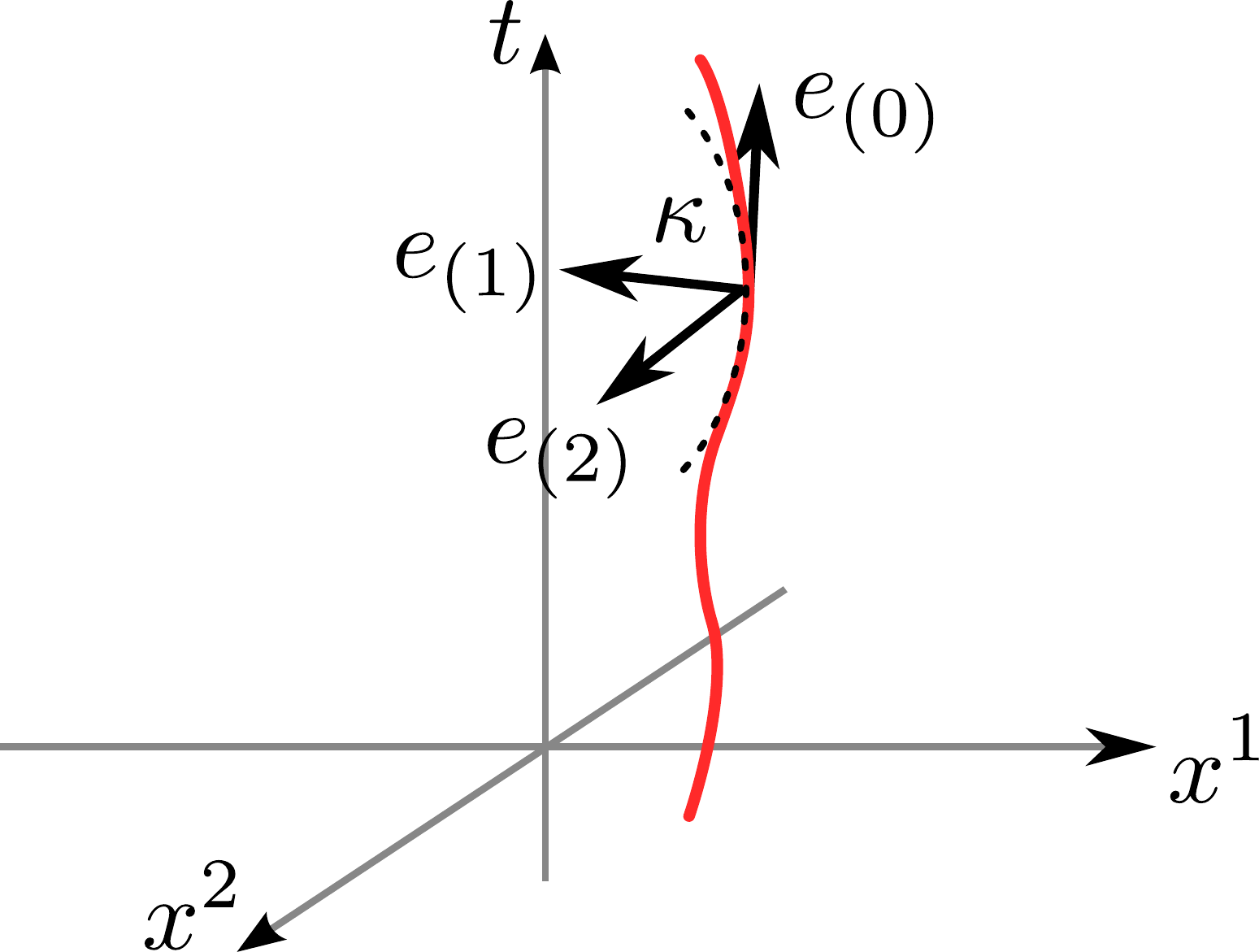}
	\end{center}
\caption{Sketch of a particle world line with the tangent vector $e_{(0)}$ and two normal vectors. The third spatial direction is not shown.
The curvature $\kappa$ determines how the tangent vector changes as one moves along the world line, $\dot e_{(1)} = \kappa e_{(1)}$.
}
\label{fig_mink}
\end{figure}

The Frenet-Serret equations are the equations of motion for the FS tetrad,
\begin{align}
\left(
\begin{matrix}
\dot e_{(0)} \\ \dot e_{(1)} \\ \dot e_{(2)} \\ \dot e_{(3)}
\end{matrix}
\right)
= 
\left(
\begin{matrix}
0 & \kappa & 0 & 0 \\ 
\kappa & 0 & \tau & 0  \\ 
0 & - \tau & 0  & \sigma \\ 
0 & 0 & -\sigma & 0
\end{matrix}
\right)
\left(
\begin{matrix}
e_{(0)} \\ e_{(1)} \\ e_{(2)} \\ e_{(3)}
\end{matrix}
\right) \,, 
\qquad \dot e_{(A)} = \sum_{B=0}^3 \Phi_{AB} e_{(B)}
\label{eq:FS-eqn}
\end{align}
determining the transport of the FS tetrad along the world line with the FS scalars determining the FS coefficient matrix $\Phi_{AB}$. See also Figure \ref{fig_mink}.

\subsection{Expansion of the World-Line in FS basis}

We can expand the world line $x(s)$ around any point
in a Taylor series in powers of $s$ around $s_0=0$
\begin{align} \label{eq:expansion-worldline}
x(s)  =  x(0) + \sum_{A=0}^3 c_A(s) \, e_{(A)}(0) 
\end{align}
which can be written in the FS basis by repeatedly employing the FS equations, Eq.~\eqref{eq:FS-eqn},
with the four coefficient functions
\begin{align}
c_0 & \simeq s + \frac{\kappa^2 s^3 }{3!} 
    + \frac{s^4}{4!} {2} \dot \kappa \kappa 
    + \frac{s^5}{5!} [ 4 \ddot \kappa \kappa + 3 \dot \kappa^2 + \kappa^4 - \kappa^2\tau^2 ] \,, \\
c_1 & \simeq  \frac{s^2}{2!} \kappa + \frac{s^3}{3!} \dot \kappa 
+ \frac{s^4}{4!} [ \ddot \kappa +\kappa (\kappa^2-\tau^2) ]
+ \frac{s^5}{5!} [\kappa^{(3)} + 3 \dot \kappa (\kappa^2-\tau^2) + 3 \kappa ( \kappa \dot \kappa - \tau \dot \tau) ]  \,, \\
c_2 & \simeq \frac{s^3}{3!} \kappa \tau  +\frac{s^4}{4!} (2\dot \kappa \tau + \kappa \dot \tau)
+ \frac{s^5}{5!} [ 3 \ddot \kappa \tau + 3\dot \kappa \dot \tau + \kappa \ddot \tau + \kappa\tau(\kappa^2-\tau^2-\sigma^2)] \,, \\
c_3 & \simeq \frac{s^4}{4!} \sigma\kappa\tau  
+ \frac{s^5}{5!} ( 3 \dot \kappa \kappa \sigma + 2 \kappa \dot \tau \sigma + \kappa\tau\dot \sigma) \,,
\end{align}
where we included all terms up to $\mathcal O(s^5)$.
All these coefficients are functions of the FS scalars and their derivatives, evaluated at $s_0=0$, i.e.~they represent the local geometry of the world line.

\begin{figure}
    \centering
    \includegraphics[width=0.6\columnwidth]{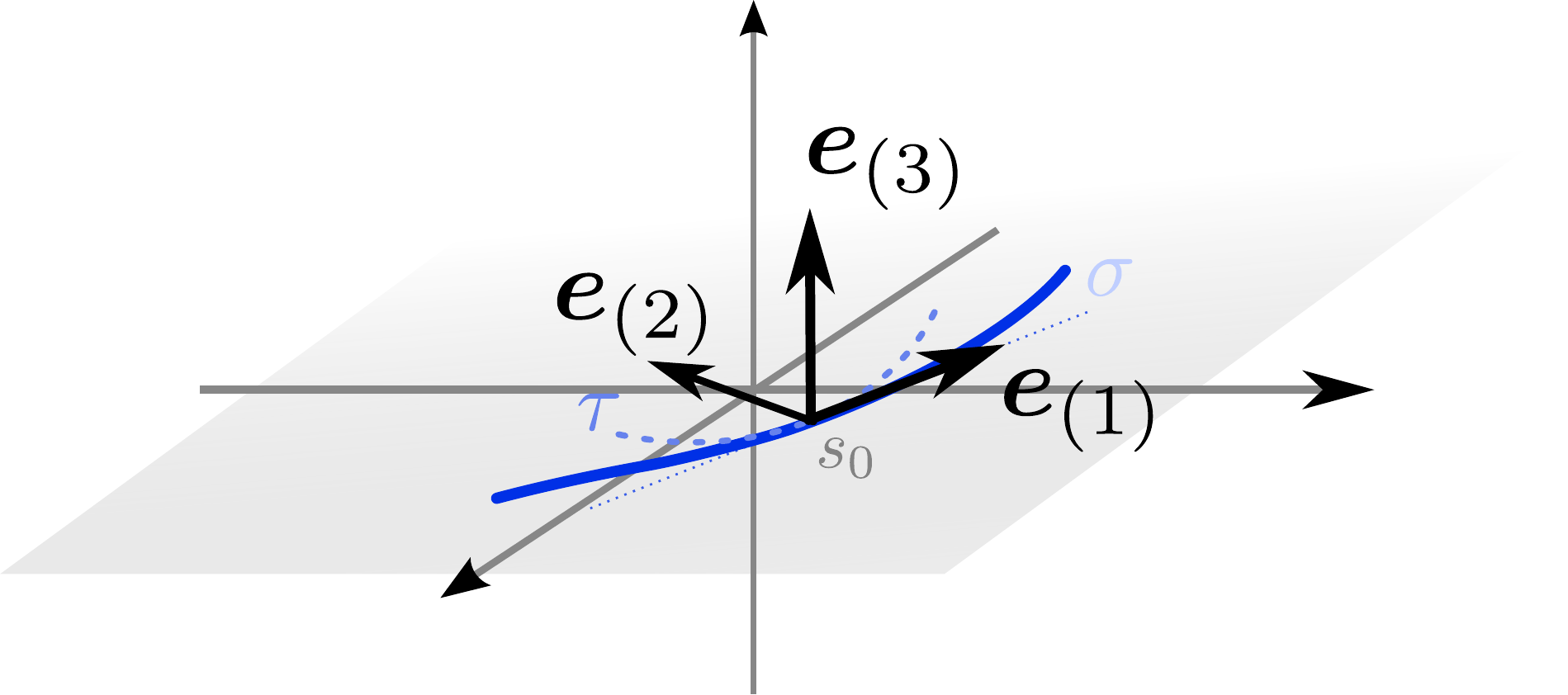}
    \caption{
    Representation of the local spatial FS basis in the instantaneous rest frame at $s_0$.
    If $\sigma=0$ the motion is restricted to a plane, which is spanned by the vectors $\vec e_{(1)}$ and $\vec e_{(2)}$, and with $\vec e_{(3)}$ being perpendicular to it.
    The reciprocal of the torsion $\tau(s_0)$ corresponds to the local curvature radius in 3-space. 
    That means for $\tau=0$ the particle moves along a straight
    line in 3-dimensional coordinate space.
    A non-vanishing value $\sigma\neq0$ determines how strongly
    the particle trajectory twists out of the plane.
    }
    \label{spatial}
\end{figure}

Based on the above expansion we can now easily discuss some special classes of motion when some of the FS scalars take special values (see \cite{Synge:ProcIrAcSc1967} for a classification for constant fields,
and Table~\ref{tab1} for more specific examples):
\begin{itemize}
\item  $\kappa=\tau=\sigma = 0$ everywhere means linear free motion $x(s) = x(0) + s u(0)$, i.e.~the particle moves with constant velocity $u^\mu(0) = e_{(0)}^\mu$.
\item The class $\kappa\neq0$ but $\sigma=\tau=0$ corresponds to a torsionless path where $c_2=c_3=0$
and the world line is $x(s) = x(0) + c_0 u(0) + c_1 e_{(1)}(0)$, i.e. the particle moves along a straight line in 3-space \cite{Formiga:AmJPhys2005}. The case of hyperbolic motion in Minkowski space-time falls into this class, i.e.~the motion of a charge in a constant electric field.
\item The class $\kappa,\tau\neq 0$, $\sigma=0$ represents the motion in a plane in 3-space, see Fig.~\ref{spatial}. It corresponds, for instance, to the motion in a constant crossed field, or a particle moving in a linearly polarized plane wave laser field. 
\end{itemize}

\section{Equations of Motion in the FS-Framework}

So far the discussion of the FS formalism was quite general.
We now specifically look into to the case of charged particles moving in strong external (laser) fields, described by
the normalized field strength tensor $f^{\mu\nu} = e F^{\mu\nu}/m$, where $e$ and $m$ are the charge and mass of the particle, respectively.

\subsection{Particles Moving in a Strong Laser Field: Lorentz Force}
\label{sect:Lorentz}

We first neglect the influence of the radiation reaction force, i.e.~we assume that the particle's motion is governed by the Lorentz force equation, $\dot u = f.u$. 

With this we immediately find
that  $\kappa^2 = u.f^2.u = \frac{e^2}{m^2} (u_\mu F^{\mu\nu} F_{\nu\lambda} u^\lambda ) = m^2\chi^2 $, i.e.~the world-line curvature is proportional to the local value of the $\chi$ parameter.
The reciprocal of the curvature represents a curvature radius
$R\sim 1/\kappa  \sim \lambdabar_C/\chi$, with $\lambdabar_C=1/m$ the reduced Compton wavelength. 
This means the onset of quantum effects for $\chi\sim 1$ relates to
the curvature radius of classical trajectories being on the order of the Compton wavelength.

We can also calculate the tetrad representation of the field strength tensor, $\mathsf F_{AB} = e_{(A)}.f.e_{(B)}$
and relate it to the FS scalars.
It can be shown that 
$$\kappa = \mathsf F_{01} 
= \frac{e}{m}\sqrt{( \gamma \vec E +\vec u \times \vec B )^2 - (\vec u\vec E)^2 }\,,$$
while $\mathsf F_{02} = \mathsf F_{03} = 0$, which is true for any particle moving in an external electromagnetic field $f$ under the influence of the Lorentz force.
In addition, relations for the derivatives of curvature can be given as
$\dot \kappa = (\dot{\mathsf F})_{01}$, $\ddot \kappa = (\ddot{\mathsf F})_{01} + \tau (\dot{\mathsf F})_{02}$,
where expressions with dots are to be understood as $(\dot{\mathsf F})_{01} \equiv e_{(0)}.\dot f . e_{(1)}$.
This means that $\dot \kappa=0$ if $f$ is constant on the world line of the particle \cite{Honig:JMathPhys1974}.
For the torsions we find the relations
$\tau = \mathsf F_{12} + \frac{1}{\kappa}(\dot{\mathsf F})_{02}$
and
$\sigma = \mathsf F_{23} + \frac{\dot \kappa}{\kappa\tau} \mathsf F_{13} 
+ \frac{2}{\tau} (\dot{\mathsf F})_{13} + \frac{1}{\kappa\tau} (\ddot{\mathsf F})_{03}$, and finally $\mathsf F_{13} = - \frac{1}{\kappa} ( \dot{ \mathsf F })_{03} $.

\begin{figure}
    \centering
    \includegraphics[width=0.9\columnwidth]{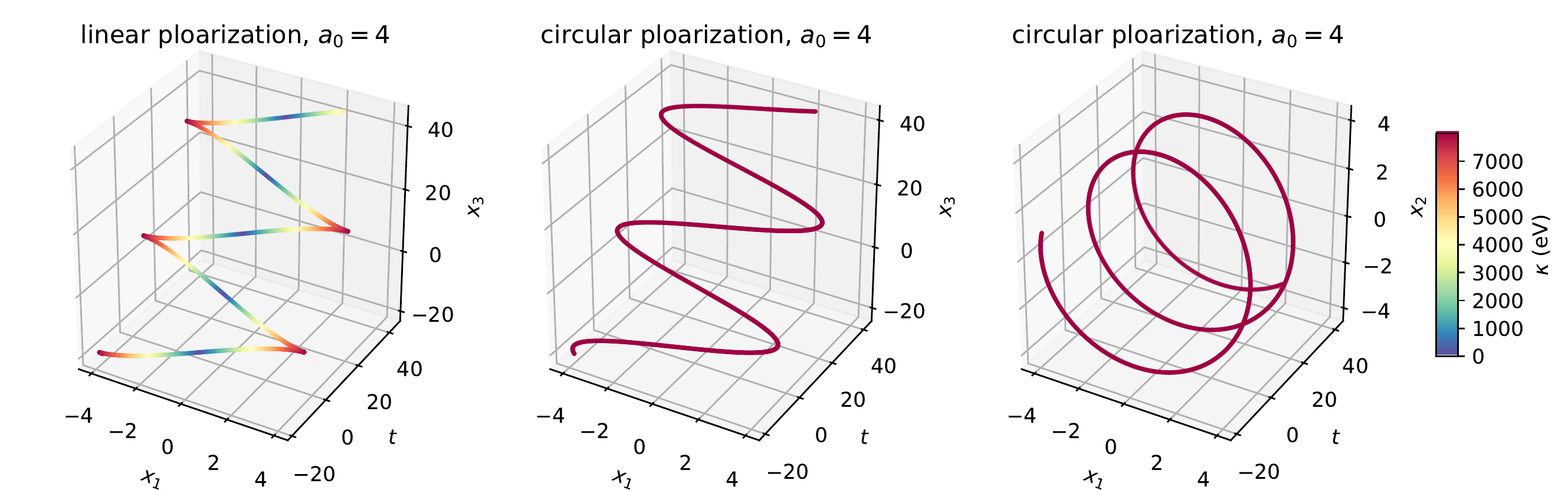}
    \caption{The world lines of an electron in a linearly polarized plane wave polarized along $x_1$ (left), and in a circularly polarized plane wave (center and right), for initial $\gamma=1000$ and $\omega=1$ eV.
    The color of the curve represents the value of the curvature $\kappa$ with blue corresponding to zero and red to the maximum value.}
    \label{fig_world_lines}
\end{figure}

For constant fields, the particle world lines are time-like helices, on which the FS scalars are constants,
and a general classification has been achieved based on the values of the field invariants and initial conditions \cite{Synge:ProcIrAcSc1967}.
In Table \ref{tab1} we collect the specific values for the curvature $\kappa$ and the torsion $\tau$ for a few important field configurations
often employed to model laser-plasma interactions, including also time-dependent fields.
Depending on the values of the FS scalars general statements about the particle motion can be made.
For instance, the curve is contained in a hyperplane if and only if the torsion $\sigma$ vanishes.
The knowledge of the FS scalars along the world line allows to reconstruct the trajectory of the particle uniquely, up to a
Poincar\'e transformation \cite{Formiga:AmJPhys2005}. 
If the initial configuration of the tetrad of FS basis vectors are known in addition further fixes the curve in space-time and just leaving the translation invariance, i.e.~the choice of an initial $x(0)$ \cite{Honig:JMathPhys1974}.
The world lines for particles moving in linearly and circularly polarized plane wave laser fields are shown in Fig.~\ref{fig_world_lines}. The color of the curve represents the curvature along those world lines. It is constant for circular polarization and oscillates in the case of linear polarization (also compare with Table \ref{tab1}).

\begin{table} 
	\begin{tabular}{p{5cm}p{3cm}l}
		\toprule[1px]
		& curvature $\kappa$ & torsion $\tau$   \\ 
		\midrule[1px]
		1D electric field$^\diamondsuit$      & $\frac{e}{m} |E_x|$ & $0$  \\
		constant magnetic field$^\heartsuit$ & $\sqrt{\gamma^2-1} \, \frac{eB_0}{m} $ & $\gamma \, \frac{e B_0}{m} $	\\				
		rotating electric field$^\spadesuit$    & 
							$\omega \, a_0 \sqrt{1+a_0^2}$
							&
						    $\omega\, (1+a_0^2)$
							 \\
		LP plane wave$^\clubsuit$   & $m b a_0 | \cos \phi |  $ & $\kappa $ \\
		CP plane wave$^\clubsuit$  & $m b a_0 = m\chi_0$ & 
		                            $m b \sqrt{1+a_0^2} $ \\
		constant crossed field$^\clubsuit$ & $m \chi_0$   & $\kappa$ \\
		\bottomrule[1px]
			\end{tabular}
			\caption{Values of the FS scalars for particle motion under the Lorentz force in specific field configurations with $\sigma=0$.
			$^\diamondsuit$ $\vec E = E_x \vec e_x$;
			$^\heartsuit$ $\vec B=B_0 \vec e_z$, zero velocity along $B$ field, valid only for $\gamma>1$ otherwise $\kappa=\tau=0$; 
			$^\spadesuit$ $E_x = E_0\cos\omega t$, $E_y= E_0\sin\omega t$, $E_z=u_z=0$,
			stationary orbits, i.e.~$\gamma=\sqrt{1+a_0^2}=const.$ 
			and $a_0=e E_0/m\omega$;
			$^\clubsuit$ laser four-momentum $k^\mu$, phase $\phi=k.x$, $b=k.p/m^2$, $\chi_0=b a_0$.
			}
\label{tab1}		
\end{table}

\subsubsection{Constant and Quasistatic Fields}

For constant fields 
further analytic results can be found also for the FS tetrad itself, and in fact have been in the literature, see e.g.~\cite{Honig:JMathPhys1974}.
The key idea is that proper time derivatives are replaced via the Lorentz force equation.
Here we find further approximations for the FS tetrad and scalars that are applicable especially for the case of high-energy particles moving in high-intensity laser pulses. The essential fact here is that the dominant parameter is $\chi\sim 1$, while the field invariants 
$ \mathcal S = \frac{1}{4m^2} f_{\mu\nu} f^{\nu\mu} = \frac{E^2-B^2}{2E_S^2}$ and
$\mathcal P = \frac{1}{4m^2}  \tilde  f_{\mu\nu} f^{\nu\mu} = \frac{\vec E\cdot \vec B}{E_S^2}$
are both much smaller than $1$ and less than $\chi$ for high-energy electrons interacting with present day high-power lasers pulses. 

While those results are exact only for homogeneous and constant fields, they could be used also as an approximation
to calculate the FS tetrad and scalars locally in quasi-static fields,
meaning that derivatives of the field strength tensor along the world line can be neglected,
$\ddot u = \frac{\ud}{\ud s} (f.u) =  \dot f . u + f^2.u  \approx f^2.u $.

From the definition of the torsions, Eqs.~\eqref{def:tau} and \eqref{def:sigma}
we find
\begin{align}
\tau^2 &= \kappa^2 - \frac{u.f^4.u}{\kappa^2} = m^2\chi^2 \left( 1 - \frac{2\mathcal S}{\chi^2} - \frac{\mathcal P^2}{\chi^4} \right)
\longrightarrow  \kappa^2  \,, \\
\sigma^2 &= \frac{\mathcal P^2}{\chi^2}
  \longrightarrow 0 \,.	
\end{align}
Typical orders of magnitude for the field invariants scale as
$\mathcal S,\mathcal P \sim E^2/E_S^2$, while the quantum nonlinearity parameter $\chi \sim  \gamma E/E_S$.
{Taking for example the record laser intensity $\unit{2\times10^{22}}{\watt\per\centi\metre^2}$ \cite{Yanovsky:OptEx2008} to estimate the typical values of the field invariants we find that
$ \mathcal S,\mathcal P \lesssim 10^{-7}$, and GeV electron colliding with such a laser pulse can reach $\chi\sim 1$. However, it should be noted that the precise values of $\mathcal S$ and $\mathcal P$ depend also on the exact field configuration and are exactly zero for plane waves, for instance. 
}
That means for particles moving with $\gamma\gg1$ through a high-intensity laser pulse
we have $\chi^2 \gg \mathcal S, \mathcal P$, and $\tau = \kappa$.

Here we used identities for the field-strength tensor $f$ and its duals $\tilde f$, such as 
$(f^4)^\mu_{\ \nu} = 2m^2 \mathcal S \, (f^2)^\mu_{\ \nu} + m^4 \mathcal P^2 \delta^\mu_\nu$ and $(f^2)^\mu_{\ \nu} - (\tilde f^2)^\mu_{\ \nu} = 2 m^2 \mathcal S \delta^\mu_\nu$.
Note that the sign of $\sigma$ is chosen such that the spatial FS triad becomes right handed, $\sigma =- \mathcal P /\chi $.
More generally, one can establish an additional relation between different FS scalars, $\kappa^2-\tau^2-\sigma^2 = 2m^2 \mathcal S $
\cite{Honig:JMathPhys1974}.

For the FS tetrad we find
\begin{align} \label{eq:e1}
e_{(1)}  &=  \frac{f.u}{\kappa} \,, \\
\label{eq:e2}
e_{(2)} &=  \frac{1}{\kappa\tau} \left[ f^2.u -\kappa^2 u \right] 
\longrightarrow \frac{f^2.u}{\kappa^2} - u \,, \\
\label{eq:e3}
e_{(3)} &= - \frac{1}{\tau}  [\tilde f.u - \frac{\beta}{\kappa^2} f.u ] 
\longrightarrow
-  \frac{\tilde f.u}{\sqrt{ u.\tilde f^2.u }}
\end{align}
The expressions on the right hand side of the long arrow are the "high-energy-approximation", $\chi^2 \gg \mathcal S,\mathcal P$ that holds for most cases when electrons interact with high-intensity laser pulses.
In this approximation $e_{(3)}$ is only approximately orthogonal to $e_{(1)}$, with their scalar products on the order of $\mathcal P/\chi^2 \ll 1$.

Before discussing in what directions those approximated basis vectors $e_{(1)}$, $e_{(2)}$, and $e_{(3)}$ in Eqs.~\eqref{eq:e1}--\eqref{eq:e3} are pointing we need to recall that we absorbed the particle's charge $e$ into the normalized field strength tensor $f$. Let us denote the sign of charge as $q = e/|e|$.
It is then straightforward to show that $e_{(1)}$ points along $q \hat {\vec E}$, 
where $\hat {\vec E}$ is the direction of the electric field in the rest frame of the particle. The vector $e_{(2)}$ points along the Poynting vector in the rest frame of the particle $\hat {\vec E}\times \hat {\vec B}$,
irrespective of the sign of the charge.
Finally, {$e_{(3)}$} points along $ - q \hat {\vec B}$ \cite{Honig:JMathPhys1974},
and $\hat {\vec E}\cdot \hat {\vec B} \ll 1$. 
This behavior is a reflection of the known fact that for ultrarelativistic particles almost all fields look locally like constant crossed fields in their instantaneous rest frames \cite{book:Landau2,Ritus:JSLR1985}.

\subsection{Classical Radiation Reaction}

Let us now investigate particle dynamics with radiation reaction (RR). If quantum stochasticity effects are not important, radiation reaction manifests itself in a smooth and continuous loss of energy and momentum of the particle, which can be modelled by adding a radiation reaction force term to the equations of motion.
The general form of classical RR equations is $\dot u = f.u + \epsilon \, P.R$,
where
$\epsilon = \frac{2\alpha}{3m}\simeq \unit{6.26\times 10^{-24}}{\second}$, with the fine structure constant $\alpha\simeq 1/137$, is the radiation reaction time scale parameter, and
$P^{\mu\nu} = g^{\mu\nu} - u^\mu u ^\nu$ is a projector onto the subspace perpendicular to $u$ in order to enforce the relativistic constraint $\dot u.u=0$.
Using the FS tetrad we can write this projector as sum over the spacelike triad $P^{\mu\nu} = - \sum_{A=1,2,3} e_{(A)}^\mu e_{(A)}^\nu$.

Different forms of the radiation reaction force exist in the literature, i.e.~the precise form of $R^\mu$ depends on RR model, see for instance the reviews \cite{Erber1961,Klepikov:PhysUspekh1985}.
For the Lorentz-Abraham-Dirac (LAD) form of the radiation reaction force, for instance, we have $R_\mathrm{LAD}^\mu = \ddot u^\mu$.
Using the FS formalism, a new RR force equation was derived in Ref.~\cite{Ringermacher:PhysLett1979} involving also the third derivative of the velocity, $\dddot u^\mu$, but later it was shown \cite{Honig:PhysLettA1981} that this equation was in principle equivalent to different equation discussed earlier by Eliezer \cite{Eliezer1947a}.

Using the FS formalism the jerk in the LAD-RR force can be expressed
$R_\mathrm{LAD} = \kappa\tau \, e_{(2)}  + \dot \kappa \, e_{(1)} + \kappa^2 \, e_{(0)}$.
Because of the projector $P$ and the orthogonality of the FS tetrad,
the time-like vector $e_{(0)}$ drops from the equation of motion,
$\dot u = f.u +\epsilon \kappa\tau e_{(2)} + \epsilon \dot \kappa e_{(1)}$.
The radiation reaction force therefore has two contributions: (i) a lateral one proportional to the torsion of the world line $\tau$, which is responsible for instance of the radiative losses due to synchrotron radiation in constant magnetic fields;
(ii) a non-stationary term proportional to the change in curvature $\dot \kappa$
that causes a radiation reaction force also in the case of vanishing lateral acceleration.
This means that in order to have any radiation reaction force acting on the particle it is required that either $\tau \neq 0 $ or $\dot \kappa \neq 0$, or both. So it becomes clear that for all torsionless paths with constant curvature, such as the hyperbolic motion of a charge in a constant electric field there is no radiation reaction force \cite{Born:AnnPhys1909}.
But a particle moving along a time-dependent electric field should indeed experience RR.

{Using the LAD equation as the equation of motion we find different relations for the FS scalars in terms of the tetrad representation of the field strength tensor. For instance, we find $(1-\epsilon \frac{\ud}{\ud s})\kappa = \mathsf F_{01}(s)$. When we try to integrate this equation assuming $\mathsf F_{01}=const.$ we find $\kappa(s) = \mathsf F_{01} + c e^{s/\epsilon}$, with a constant $c$. 
}

It is well known that the  LAD equation, as it is involving the derivative of acceleration $\ddot u$, can lead to unphysical solutions in the form of runaways with
a non-perturbative dependence on the electromagnetic coupling $\alpha$ in the form
$e^{s/\epsilon} \sim  1 + \mathcal O(\alpha^{-1})$ \cite{Bhabha:PR1946}.
The above solution for $\kappa$ represents such a runaway with the same non-perturbative dependence on $\epsilon$ respectively $\alpha$.
Landau and Lifshitz proposed to reduce the order by iterating the equation treating $\epsilon$ as a small parameter, and keeping only
terms linear in $\epsilon$ \cite{book:Landau2}.
In the FS formalism that means we can replace all FS scalars and basis vectors multiplying $\epsilon$ by their Lorentz force equivalents discussed in Section \ref{sect:Lorentz}.
Applying this to the above equation for $\kappa$ we find
$\kappa(s) = \mathsf F_{01}(s) + \epsilon \dot \kappa_\mathrm{Lorentz}$ up to linear order in $\epsilon$, which is free of runaway solutions.
As another interesting fact we find that $\mathsf F_{02} = - \epsilon\kappa\tau$ for a particle moving under the radiation reaction force, in contrast to the Lorentz force case where $\mathsf F_{02}=0$ was found.
{This shows that the vector $e_{(2)}$ is not orthogonal to the rest frame electric field $f.e_{(0)}$
when RR is taken into account. Since the deviation is proportional to the lateral acceleration $\tau$,
it would be interesting to investigate further how this relates to the
concept of the radiation-free-direction.
It was argued in Ref.~\cite{Gonoskov:PhysPlas2018} that radiative losses lead to a tendency of charged particles to align their propagation direction along the local radiation-free-direction, for which the lateral acceleration is minimized.
}

\subsection{Classical Spin Precession in Strong Laser Fields: T-BMT Equation in the FS Formalism}

The Thomas-Bargman-Michel-Telegdi (T-BMT) equation \cite{Bargmann:PRL1959} that describes the classical precession of the spin-polarization 4-vector $S^\mu$ reads
\begin{align}
\frac{\ud S^\mu}{\ud s} = \frac{g}{2} \left[ f^{\mu\nu} S_\nu + u^\mu (S.f.u)\right] - u^\mu (S.\dot u) \,,
\end{align}
with the $g$-factor of the electron.
Because $S^\mu$ is a space-like vector that is perpendicular to the 4-velocity ($S.u=S.e_{(0)}=0$) it seems natural to expand $S$ in the space-like FS-triad,
\begin{align}
S^\mu = \sum_{A=1,2,3} S_A e^\mu_{(A)}  \,, 
\end{align}
upon which the T-BMT equation becomes 
\begin{align}
\sum_{A=1,2,3} \dot S_A e_{(A)} 
		= \sum_{A=1,2,3} \frac{g}{2} [ f.e_{(A)} S_A + u \, S_A (e_{(A)}.f.u) ]	-   S_A (e_{(A)} . \dot u)\, u - S_A \dot e_{(A)} \,.
\end{align}
Using the orthogonalyity of the FS tetrad $e_{(A)}.e_{(B)} = g_{AB} 
$ we can derive
 equations for the coefficients $S_A$,
\begin{align}
- \dot S_B &= \sum_{A=1,2,3} \frac{g}{2} ( e_{(B)}.f.e_{(A)}) S_A - S_A (e_{(B)}. \dot e_{(A)}) \,.
\end{align}
After using the FS equations, Eq.~\eqref{eq:FS-eqn}, the T-BMT equation can be cast into
\begin{align} \label{eq:BMT-final}
\dot S_B &= \sum_{A=1,2,3} \left(  \Phi_{BA}  - \frac{g}{2} \mathsf F_{BA} \right) S_A = \sum_{A=1,2,3} \mathsf H_{BA} S_A \,,
\end{align}
where $\mathsf F_{BA} = e_{(B)}.f.e_{(A)}$ is the FS representation of the field strength tensor and we made use of the symmetry properties of both $\mathsf F$ and $\Phi$. ($\sf F$ is antisymmetric in its tetrad indices. For more discussion on the expression of the Faraday tensor in terms of the FS tetrad see also Ref.~\cite{Caltenco:CzechJP2002}).
Because of the antisymmetry, the matrix $\mathsf H$ is orthogonal and the length of the tetrad representation of the spin vector $S$ is conserved.
In the space-like part of the $\Phi_{AB}$ only the two torsions $\tau$ and $\sigma$ of the world line appear,
but not the curvature $\kappa$. The 2nd term in the brackets in \eqref{eq:BMT-final} is due to the direct action of the field, while the first part is due to the geometry of the world-line.

Using the results from Section \ref{sect:Lorentz} it is straightforward to show that in a constant field the non-vanishing components of $\mathsf F$ are
$\mathsf F_{12} = \tau$ and $\mathsf F_{23} = \sigma$. Hence, we find that for 
$g=2$ (i.e.~no anomalous magnetic moment of the electron) the spin-polarization vector does not precess with respect to the co-transported Frenet-Serret basis, i.e.~$\dot S_A=0$.
Honig showed this by explicitly going to the instantaneous particle rest frame \cite{Honig:JMathPhys1974}.

This is an important result for the application of LCFA scattering rates in Monte Carlo simulations for quantum radiation
reaction with spin-polarized particles. The scattering rates 
are usually calculated using S-matrix theory using the {locally constant field approximation (LCFA)},
and therefore depend on the asymptotic spin-properties. Contrary, in the simulation one tracks the classical evolution of the spin via the T-BMT equation and calculates the local polarization vector inside the field at finite time. In order to connect the two one needs to identify constant of motion in a constant crossed field. {The above result provides exactly that:}
The expansion coefficients of the spin-vector in the local FS basis $S_A$ are constant in a constant field and can therefore be 
reinterpreted as the asymptotic polarization properties entering the polarization dependent scattering rates.
For instance, the basis vectors used in the calculation of the scattering rates in \cite{Seipt:PRA2018} are the asymptotic limits of the
FS basis vectors in their approximated form given in Eqs.~\eqref{eq:e1}--\eqref{eq:e3}.

\section{Scattering}

For strong-field QED scattering processes like non-linear Compton scattering the event probabilities are expressed in terms
of mod squared Furry picture S-matrix elements \cite{Ritus:JSLR1985,Seipt:PRA2015,Seipt:PRL2017,Ilderton:PRD2018,Kharin:PRL2018}.
These expressions typically involve integrals over the phase space of the outgoing particles and
two laser phase variables, corresponding to the the S-matrix ($\phi$) and its complex conjugate ($\phi'$) \cite{Dinu:PRA2013,Ilderton:2018}.
One of the key elements, determining the phase exponent of the scattering probability, is the (normalized squared) Kibble mass $\mu = \langle u_\nu \rangle \langle u^\nu \rangle$, which relates to the mass shell condition of the averaged kinetic electron momentum,
and where
\begin{align}
\langle u^\mu \rangle = \langle u^\mu \rangle (\varphi,\theta) = \frac{1}{\theta} \int_{{\varphi -\theta/2}}^{\varphi+\theta/2} \! \ud \phi \, u^\mu(\phi) \,,
\end{align}
denotes a laser phase average over a window of size $\theta = \phi'-\phi$  around the midpoint $\varphi = (\phi+\phi)/2$ \cite{Kibble:NPB1975}.

In a plane wave laser field with four wave-vector $k$, which depends only on the phase variable $\phi=k.x$, the particle's proper time $s$ is proportional to the laser phase \cite{Ritus:JSLR1985},
$\phi =mbs+\phi_0$ with the quantum energy parameter $b=k.p/m^2$ which equals the laser frequency in the electron rest frame in units of the electron rest mass, and 
with $p$ as the electron four-momentum. 
This implies that we can always replace the phase by proper time and we can apply the FS formalism.

With this we can calculate the average of the velocity as 
$\langle u^\mu \rangle =  [ x^\mu(s/2) - x^\mu(-s/2)]/s = \sum_{A=0}^3  g_A e^\mu_{(A)}\,,$
and by using the expansion of the world line Eq.~\eqref{eq:expansion-worldline}, the expansion coefficients of $\langle u\rangle$ read
\begin{align}
g_0 &= 1 + \frac{\theta^2  \kappa^2 }{3!2^2(mb)^2}  + \frac{\theta^4}{5!2^4(mb)^4} [ 4 \ddot \kappa \kappa + 3 \dot \kappa^2 + \kappa^2(  \kappa^2- \tau^2) ] \,,\\
g_1 &= \frac{\theta^2}{3!2^2(mb)^2} \dot \kappa 
+ \frac{\theta^4}{5!2^4(mb)^4} [\dddot \kappa + 3 \dot \kappa (\kappa^2-\tau^2) + 3 \kappa ( \kappa \dot \kappa - \tau \dot \tau) ] \,,\\
g_2 &= \frac{\theta^2}{3!2^2(mb)^2} \kappa \tau 
+ \frac{\theta^4}{5!2^4(mb)^4} [3 \ddot \kappa \tau + 3\dot \kappa \dot \tau + \kappa \ddot \tau + \kappa\tau(\kappa^2-\tau^2-\sigma^2)]  \,, \\
g_3  &= \frac{\theta^4}{5!2^4(mb)^4} ( 3 \dot \kappa \kappa \sigma + 2 \kappa \dot \tau \sigma + \kappa\tau\dot \sigma) \,.
\end{align}
The $g_A$ and hence the FS scalars are evaluated at the midpoint of the averaging window $\varphi$, e.g.~$\kappa=\kappa(\varphi)$.

With these results we can write the approximation for the Kibble mass $\mu = \langle u\rangle^2$. Because the FS tetrad is an orthonormal basis,
$e_{(A)} . e_{(B)} = g_{AB}$, we have
\begin{align}
\mu \simeq \sum_{A=0}^3 e_A^2 g_A^2  = g_0^2 - g_1^2 - g_2^2 \,, 
\end{align}
because $g_3^2 = \mathcal O(\theta^8)$ and we are keeping only terms up to $\theta^4$. 
Finally,
\begin{align}
\label{eq:Kibblemass-expansion-FS}
\mu \simeq 1 + \frac{\theta^2}{12(mb)^2} \kappa^2 + \frac{\theta^4}{720(mb)^4} [ \dot \kappa^2 + 3 \kappa\ddot \kappa  + 2 \kappa^2 (\kappa^2-\tau^2) ] \,.
\end{align}

In the lowest non-trivial order (in $\theta$) the Kibble mass depends only on the local curvature of the world line, i.e.~the local $\chi$-factor of the electron, $\kappa/mb = \chi(\varphi) /b$.
The world line is approximated locally as an arc with constant curvature.
This is of course the basis for the locally constant field approximation (LCFA) widely used to calculate photon emission rates 
for electron interaction with high-intensity laser fields, i.e. for the simulation of quantum radiation reaction effects.

The next-to-leading order contains the derivatives of the curvature
and the torsion $\tau$ of the world line as two different effects \cite{Khokonov:PRL2002,Ilderton:2018}.
It is worth noting that both in \eqref{eq:Kibblemass-expansion-FS}
	and in the terms comprising the classical RR the second torsion $\sigma$ does not appear at all.

Upon specifying a plane wave field $f^{\mu\nu}(\phi) = a_0 \sum_{j=1,2}  h_j(\phi) (k^\mu \varepsilon^\nu_j - k^\nu \varepsilon^\mu_j)$ , with $a_0$ the normalized laser vector potential, we find that 
\begin{align}
\dot \kappa^2 + 3 \kappa\ddot \kappa + 2 \kappa^2 (\kappa^2 -  \tau^2 )
&= 
 m^4 b^4 a_0^2 \: \sum_j \left[ h_j' h_j' + 3 h_j'' h_j  \right] \,,
\end{align}
with a prime denoting a phase derivative. 
This form of the correction to the LCFA directly in terms of field gradients was derived, e.g. in Refs.~\cite{Baier:JETP1981,Ilderton:2018}.
These corrections were used to construct improved photon emission rates that go beyond the LCFA by including field gradient effects, termed \LCFAp \cite{Ilderton:2018}.
See also \cite{DiPiazza:PRA2018} for a different approach to improve the LCFA scattering rates.

The FS scalars in \eqref{eq:Kibblemass-expansion-FS} can also be expressed in terms of proper time derivatives of the particle four-velocity, yielding
\begin{align}
\mu \simeq  1 - \frac{s^2}{12} (\dot  u.\dot u) - \frac{s^4}{720} \left[ (\ddot u. \ddot u) + 3 (\dot u. \dddot u) \right] \,.
\end{align}
which could be used to implement \LCFAp scattering rates into the QED modules of particle in cell codes
\cite{Ridgers:JCompPhys2014,Vranic:NJP2016}
where the fields are not necessarily plane waves.

\section{Experimental distinction of classical and quantum RR}

In the following we present a numerical simulation where we implemented the improved \LCFAp photon emission rates \cite{Ilderton:2018}
using the Monte Carlo algorithm outlined in Ref.~\cite{Duclous:PPCF2011,Ridgers:JCompPhys2014}. 
Two independent CLF experiments recently verified the occurrence of radiation reaction effects in high-power laser-electron-beam scattering experiments \cite{Cole:PRX2018,Poder:PRX2018}. However, the stochasticity of the quantum RR could not yet be verified unequivocally. The main reason being the electron beams that were produced using LWFA 
did not show prominent quasi-monoenergetic features which could be used to gauge the influence of RR effects.

\begin{figure}[ht]
	\includegraphics[width=0.7\columnwidth]{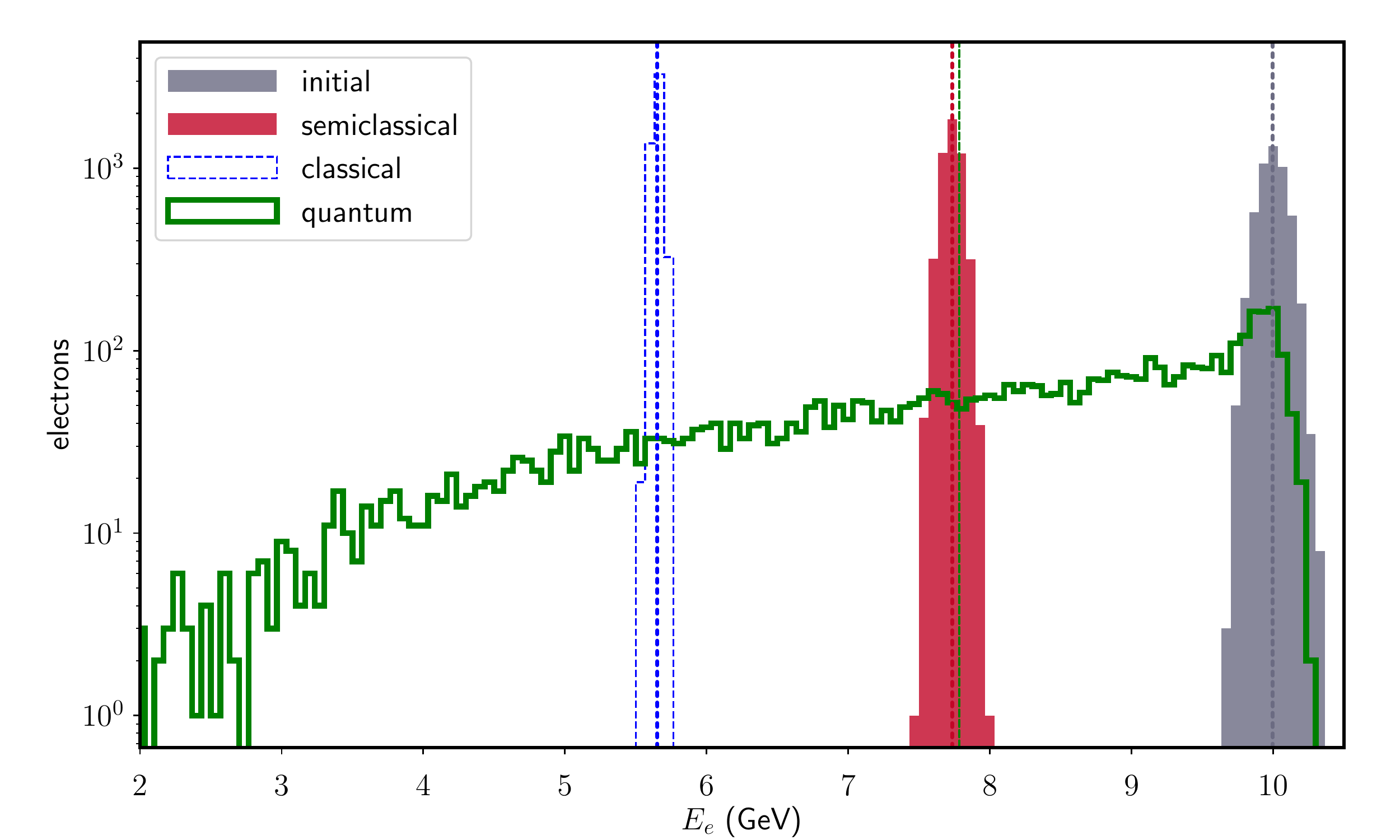}
	\caption{Energy spectra of electrons after undergoing interaction with a high intensity laser pulse. Initial conditions: 10 GeV mean energy and 1\% energy spread. Laser: $a_0=5$, $T = 44$ fs FWHM, $I = \unit{0.5\times 10^{20} }{\watt\per\centi\metre^2}$. {The vertical lines mark the average energies for each distribution.}}
	\label{fig1}
\end{figure}

Using a conventional RF accelerator instead would allow to better control the initial electron beam to be used in the scattering experiment. An experiment where a conventional electron beam was brought into collision with a (by today's standards moderately) intense laser beam was the very successful SLAC E-144 \cite{Bamber:PRD1999}, where both non-linear Comtpon scattering and Breit-Wheeler pair production were investigated.
With an upgraded PW laser system at SLAC to allow for higher laser intensity a repetition of this experiment could allow to distinguish the classical from the quantum RR regimes.
A simulation of such an experiment is shown in Fig.~\ref{fig1}.

Some of the most important features to make such an experiment successful include:
(i) Quasi monoenergetic initial electron beam energy, in our simulations we take 1 percent relative energy spread. 
(ii) $\chi\lesssim 1$, in order to enhance quantum stochasticity effects which are suppressed for $\chi\ll1$ \cite{Blackburn:PRL2014,Ridgers:JPP2017,Niel2017}, but keep it small enough to mitigate non-linear Breit-Wheeler pair production by the emitted $\gamma$-rays \cite{DiPiazza:RevModPhys2012}. 
The $\chi_\gamma$ factor of the photons is always smaller than the $\chi$ factor of the electrons, and pair production is suppressed for $\chi_\gamma \lesssim 1$.
(iii) A large laser focal spot in order to mitigate non-ideal effects such as ponderomotive scattering as well as focus averaging effects.
A 1 PW laser focused down to a spot size of $w_0 = \unit{45}{\micro\metre}$ will generate a peak intensity of
$\unit{0.5\times 10^{20}}{\watt\per\centi\metre^2}$, or $a_0=5$. 
Colliding this laser with a 10 GeV electron beam provides a peak value of $\chi = 0.6$. This is large enough for seeing quantum effects, but sufficiently small to mitigate the possibility of pair production by the emitted photons \cite{Niel2017}.

A value of $a_0=5$ is quite low for the applicability of the locally constant field approximation for the scattering rates, which becomes better as $a_0\to \infty$.
{The LCFA is known to overestimate the number of emitted photons and also the emitted power, especially for small values of $a_0$ \cite{Blackburn:PhysPlas2018}.}
We therefore employ here improved photon emission rates, which have been derived in \cite{Ilderton:2018}. 
These \LCFAp rates take into account field gradient effects, and as has been discussed above this can be seen as corrections due to a non-constant curvature of the particle world-line and its torsion.

{The laser is modeled here as a pulsed plane wave with FWHM duration $44$ fs,
	and with the electrons initially counterpropagating. The classical electron dynamics is described by solving the Lorentz force equation for the four velocity $u^\mu(s)$ in proper time using a fourth-order Runge Kutta algorithm. Numerical convergence is tested by verifying that $u.u= \gamma^2 - \vec u^2 = 1$.
	The QED emission model follows closely the one described in Refs.~\cite{Duclous:PPCF2011,Ridgers:JCompPhys2014}.
	At the beginning of the simulation, and after every photon emission, each electron is assigned a final optical depth and the current optical depth is set to zero.
	 For each timestep the probability $\Delta \mathbb P$ for photon emission during that step is calculated using the \LCFAp photon emission rates \cite{Ilderton:2018}, and added to the current optical depth until the final optical depth is reached, upon which the energy of the photon is sampled from the normalized photon energy spectrum and the momentum of the photon is subtracted from the electron by assuming the photon is emitted parallel to the instantaneous electron velocity $\vec u$. A total of $10^3$ particles have been simulated to sample the electron distributions.}

The final electron energy distribution in the quantum calculation (see Fig.~\ref{fig1}) becomes very broad due to the stochasticity in photon emission. This has to be compared to the classical and semiclassical radiation reaction models which predict a narrowing of the initial energy spread.
The notion semiclassical refers to a model where the radiation reaction terms in the classical Landau-Lifshitz equation, are modified by a $\chi$ dependent Gaunt factor
{$\epsilon \to \epsilon g(\chi)$, with
$g(\chi) =  ( 1 + 3.72 ( 1 + \chi ) \log( 1 + 2.34  \chi ) + 2.80  \chi^2 )^{-2/3}$},
which takes into account the reduced emission due to quantum effects, but not the stochasticity \cite{Kirk:PPCF2009,Thomas:PRX2012,DelSorbo:PPCF2018}.
The final electron spectra shown in Fig.~\ref{fig1} clearly
shows that the mean energy (red vertical line) of the semiclassical model agrees well with
the full quantum calculation (green vertical line), and differs significantly from the classical model
{which predicts a much lower mean\textbf{} final electron energy.}
Moreover, it demonstrates the spectral narrowing for the (semi)classical models and the spreading for the stochastic quantum model of radiation reaction.

\section{Summary}

In this paper we applied the Frenet-Serret formalism to the
dynamics of charged particles in high-intensity laser fields.
We find approximate relations for the Frenet-Serret scalars and basis vectors relevant for high-intensity laser particle interactions.
We find that the onset of quantum effects at $\chi\sim 1$ relates to
the curvature radius $\kappa^{-1}$ of classical trajectories being on the
order of the Compton wavelength.
We discuss both classical and quantum radiation reaction effects and 
represent the classical spin-precession in terms of the co-moving space-like
Frenet-Serret basis. This furnishes a covariant proof that in a constant field, and for $g=2$, the spin four-vector does not precess with regard to the Frenet-Serret basis.
We comment on the derivation of the photon emission rate in strong-field QED beyond the locally constant field approximation,
which is used in Monte Carlo simulations of quantum radiation reaction.
We conclude by discussing a possible experiment to distinguish between classical and quantum mechanical models of radiation reaction.

\subsection*{Acknowledgements}

D.~S.~acknowledges fruitful discussions with A.~Ilderton and B.~King. This work was funded in part by the US ARO grant no.~W911NF-16-1-0044.

%

\end{document}